\documentclass[aps,pra,showpacs, twocolumn,amsmath,amssymb,tightenlines,epsfig,floatfix]{revtex4-1}
\usepackage{graphicx}
\usepackage{dcolumn}	
\usepackage{bm}			
\usepackage{amsfonts}
\usepackage{xspace}
\usepackage{color}
\usepackage{epstopdf}

\usepackage{multirow}
\usepackage{array}

\begin{document}

\newcommand{\psihat}{\ensuremath{\hat{\psi}}\xspace}
\newcommand{\psihatd}{\ensuremath{\hat{\psi}^{\dagger}}\xspace}
\newcommand{\ahat}{\ensuremath{\hat{a}}\xspace}
\newcommand{\Ham}{\ensuremath{\mathcal{H}}\xspace}
\newcommand{\ahatd}{\ensuremath{\hat{a}^{\dagger}}\xspace}
\newcommand{\bhat}{\ensuremath{\hat{b}}\xspace}
\newcommand{\bhatd}{\ensuremath{\hat{b}^{\dagger}}\xspace}
\newcommand{\chat}{\ensuremath{\hat{c}}\xspace}
\newcommand{\chatd}{\ensuremath{\hat{c}^{\dagger}}\xspace}
\newcommand{\boldr}{\ensuremath{\mathbf{r}}\xspace}
\newcommand{\dr}{\ensuremath{\,d^3\mathbf{r}}\xspace}
\newcommand{\dk}{\ensuremath{\,d^3\mathbf{k}}\xspace}
\newcommand{\etal}{\emph{et al.\/}\xspace}
\newcommand{\ie}{i.e.\:}
\newcommand{\eq}[1]{Eq.\,(\ref{#1})\xspace}
\newcommand{\eqs}[1]{Eqs.\,(\ref{#1})\xspace}
\newcommand{\fig}[1]{Figure\,(\ref{#1})\xspace}
\newcommand{\abs}[1]{\left| #1 \right|} 
\newcommand{\proj}[2]{\left| #1 \rangle\langle #2\right| \xspace} 
\newcommand{\Qhat}{\ensuremath{\hat{Q}}\xspace}
\newcommand{\Qhatd}{\ensuremath{\hat{Q}^\dag}\xspace}
\newcommand{\phihatd}{\ensuremath{\hat{\phi}^{\dagger}}\xspace}
\newcommand{\phihat}{\ensuremath{\hat{\phi}}\xspace}
\newcommand{\boldk}{\ensuremath{\mathbf{k}}\xspace}
\newcommand{\boldp}{\ensuremath{\mathbf{p}}\xspace}
\newcommand{\boldsigma}{\ensuremath{\boldsymbol\sigma}\xspace}
\newcommand{\boldalpha}{\ensuremath{\boldsymbol\alpha}\xspace}
\newcommand{\grad}{\ensuremath{\boldsymbol\nabla}\xspace}
\newcommand{\parti}[2]{\frac{ \partial #1}{\partial #2} \xspace}
 \newcommand{\vs}[1]{\ensuremath{\boldsymbol{#1}}\xspace}
\renewcommand{\v}[1]{\ensuremath{\mathbf{#1}}\xspace}
\newcommand{\Psihat}{\ensuremath{\hat{\Psi}}\xspace}
\newcommand{\Psihatd}{\ensuremath{\hat{\Psi}^{\dagger}}\xspace}
\newcommand{\Vhatd}{\ensuremath{\hat{V}^{\dagger}}\xspace}
\newcommand{\Xhat}{\ensuremath{\hat{X}}\xspace}
\newcommand{\Xhatd}{\ensuremath{\hat{X}^{\dag}}\xspace}
\newcommand{\Yhat}{\ensuremath{\hat{Y}}\xspace}
\newcommand{\Yhatd}{\ensuremath{\hat{Y}^{\dag}}
\xspace}
\newcommand{\ddt}{\ensuremath{\frac{d}{dt}}
\xspace}
\newcommand{\nset}{\ensuremath{n_1, n_2,\dots, n_k}
\xspace}
\newcommand{\sah}[1]{{\textcolor{red}{#1}}}
\newcommand{\jhat}{\ensuremath{\hat{J}}
\xspace}

\title{Generation of Atom-Light Entanglement in an Optical Cavity for Quantum Enhanced Atom-Interferometery}
\author{Simon A. Haine}
\email{haine@physics.uq.edu.au}
\affiliation{School of Mathematics and Physics, University of Queensland, Brisbane, QLD, 4072, Australia}
\author{Wing Yung Sarah Lau} 
\affiliation{Australian Research Council Centre for Engineered Quantum Systems, and Australian Research Council Centre for Quantum Computing and Communication Technology, School of Mathematics and Physics, University of Queensland, Brisbane, QLD, 4072, Australia}

\begin{abstract}
We theoretically investigate the generation of atom-light entanglement via Raman superradiance in an optical cavity, and show how this can be used to enhance the sensitivity of atom interferometry. We model a realistic optical cavity, and show that by careful temporal shaping of the optical local oscillator used to measure the light emitted from the cavity, information in the optical mode can be combined with the signal from the atom interferometer to reduce the quantum noise, and thus increase the sensitivity. It was found in Phys.~Rev.~Lett.~{\bf 110}, 053002 (2013) that an atomic `\emph{seed}' was required in order to reduce spontaneous emission and allow for single mode behaviour of the device. In this paper we find that the optical cavity reduces the need for an atomic seed, which allows for stronger atom-light correlations and a greater level of quantum enhancement. 
\end{abstract}


\maketitle
\section{Introduction}
Inertial sensors based on atom interferometers have the potential to provide state-of-the-art sensitivity for a range of scientific applications \cite{Cronin:2009, Peters:1999, McGuirk:2002, Fixler:2007}. Although most state-of-the-art atom interferometers currently utilize laser cooled thermal atoms, there are some benefits to using Bose-Einstein-condensed atoms, as they provide improved visibility in configurations which require complex manipulation of the motional state such as high momentum transfer beam splitters \cite{Debs:2011, Altin:2011, Szigeti:2012, Hardman:2014}. Typically, these devices utilise uncorrelated sources of atoms, so cannot resolve phase shifts smaller than $\frac{1}{\sqrt{N_t}}$, where $N_t$ is the total number of particles \cite{Dowling:1998}. This is known as the standard quantum limit (SQL). There is recently considerable interest in the development of \emph{quantum-enhanced} atom interferometry, which allows for sensitivities beyond the SQL. Such schemes rely on the use of entangled many-body quantum states, which can be generated via atomic interactions \cite{Esteve:2008, Gross:2010, Riedel:2010, Hamley:2012, Berrada:2013, Lucke:2014, Muessel:2015}, or atom-light interactions \cite{Hald:1999, Kuzmich:2000, Appel:2009, Leroux:2010, Schleier-Smith:2010b, Chen:2011, Sewell:2012, Haas:2014, McConnell:2015}. However, it was found in \cite{Anderson:2009, Haine:2009, Haine:2011, Haine:2014} that in some circumstances that the strong nonlinear atomic interactions required for entanglement generation can adversely affect the ability to mode-match the two arms of the interferometer, and therefore diminish the interferometer signal. It has recently been shown that Raman superradiance \cite{Moore:2000, Schneble:2004, Yoshikawa:2004, Cola:2004, Wang:2005, Uys:2007, Hilliard:2008} can be used used to enhance the sensitivity of atom interferometry \cite{Haine:2013}.  This approach may have some advantages over other approaches, as the modes that this process generates are automatically mode-matched to the Raman transitions used as beam-splitters in atom-interferometer-based inertial sensors. In \cite{Haine:2013}, it was found that a small `seed' of atoms was required in each atomic mode in order to suppress spontaneous emission and produce two modes with well defined momentum. However, it was found that this seed reduced the level of quantum correlations, reducing the amount of possible quantum enhancement. 

In this paper we consider the use of an optical cavity to enhance the coupling into one particular mode. As the optical cavity allows for the light to interact with the atoms for longer, this creates a greater level of Bose-enhancement into one particular atomic mode, allowing for a smaller atomic seed, and thus stronger atom-light quantum correlations. We investigate a realistic cavity and show how this process can be used to generate atom-light entanglement, and how these correlations can be used to enhance the sensitivity of atom interferometry.  

\section{Scheme}
Our scheme is described in Fig.~\ref{scheme1}, and can be separated into two stages: 
\begin{enumerate}
\item A state preparation stage, where the super-radiance drives atomic population dynamics and creates atom-light correlations;
\item A measurement stage, where this state is used as the input to an atom interferometer which is used to estimate some physical quantity.  
\end{enumerate}
Briefly, a condensate consisting of 3-level atoms (two non-degenerate hyperfine ground states $|1\rangle$ and $|2\rangle$, and an excited state $|3\rangle$) with the entire population initially in state $|1\rangle$  is placed in an optical cavity. The atoms are then optically driven by a classical pump field detuned from the $|1\rangle \rightarrow |3\rangle$ transition by an amount $\Delta_1$. Emission of a photon into the cavity mode (detuned from the $|2\rangle \rightarrow |3\rangle$ transition by frequency $\Delta_2$) results in the creation of a state $|2\rangle$ atom.  We choose the frequency of the driving field and the cavity mode such that they achieve two-photon resonance for the creation of a state $|2\rangle$ atom. As the creation of a cavity mode photon is associated with the creation of a state $|2\rangle$ atom, we expect correlations between these two modes. After a small number of state $|2\rangle$ atoms have been created, the driving is turned off and the light is allowed to leak out of the cavity, and measured via homodyne detection. The state $|2\rangle$ atoms are then combined with the remaining state $|1\rangle$ atoms as the input to a Mach-Zehnder (MZ) interferometer, formed by coherently coupling these two modes via optical Raman transitions. The atomic population difference is then measured and combined with information from the optical homodyne measurement in order to extract the atomic phase shift. We will now discuss the theoretical model for each stage in detail:

\begin{figure}[h]
\includegraphics[width=0.9\columnwidth]{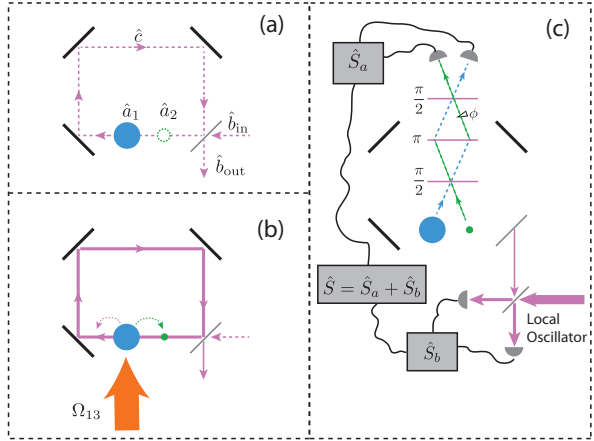}
\caption{(Color online). Scheme for quantum-enhanced atom-interferometry. (a) and (b): State preparation. State $|1\rangle$ atoms (annihilation operator $\ahat_1$) are driven by a classical pump field (Rabi-Frequency $\Omega_{13}$), leading to the creation of a small number of state $|2\rangle$ atoms (annihilation operator $\ahat_2$), and photons in the cavity mode (annihilation operator $\chat$). (c): Measurement: The classical pump is turned off, and the photons leaking out of the cavity are measured via homodyne detection, and the two atomic modes ($\ahat_1$ and $\ahat_2$) are used as the input to a Mach-Zehnder Interferometer. 
}
\label{scheme1}
\end{figure}

\subsection{State preparation}
\begin{figure}[h]
\includegraphics[width=0.5\columnwidth]{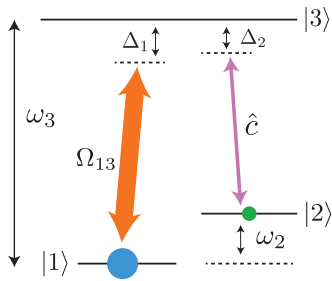}
\caption{(Color online). Energy level scheme for a three-level Raman transition comprising two nondegenerate hyperfine ground states ($|1\rangle$ and $|2\rangle$). The BEC is initially formed in state $|1\rangle$, and populations is transferred to $|2\rangle$ via the absorption of a photon from from the classical pump beam (Rabi frequency $\Omega_{13}$, detuned from the excited state by $\Delta_1$) and the emission of a photon into the cavity mode $\chat$ (detuned from the excited state by $\Delta_2$).
}
\label{levels}
\end{figure}

Our model has previously been described in \cite{Haine:2013}, and is summarised in Fig.~\ref{levels}. We begin with a BEC of 3-level atoms: Two non-degenerate hyperfine ground states $|1\rangle$ and $|2\rangle$, and an excited state $|3\rangle$. All the atoms are initially in state $|1\rangle$. The optical cavity mode (annihilation operator $\chat$) is detuned from the $|2\rangle \rightarrow |3\rangle$ transition by $\Delta_2$. The $|1\rangle \rightarrow |3\rangle$ transition is driven by a strong pump laser of Rabi frequency $\Omega_{13}$ and detuning $\Delta_1$.  After adiabatically eliminating the excited state as in \cite{Sinatra:1995, Moore:1999, Szigeti:2014b}, the Hamiltonian describing the system is
\begin{eqnarray}
\Ham &=& \sum_{j=1,2} \int \psihatd_j(\boldr) \hat{H}_j \psihat_j(\boldr) \dr \nonumber +  \hbar \left(\omega_3-\Delta_2\right) \chatd\chat \\
&+& g_c \frac{\Omega_{13}}{\Delta_1}\int\left(  \Lambda(\boldr, t)\chatd \psihat_1(\boldr)\psihatd_2(\boldr) + {\rm h.c.} \right) \dr \nonumber \, , \\
\end{eqnarray}
where
\begin{equation}
\Lambda(\boldr, t) = e^{i \left((\boldk_1 -\boldk_2)\cdot \boldr - (\omega_3-\Delta_1)t \right)} \, ,
\end{equation}
and $\psihat_j(\boldr)$ annihilates a state $|j\rangle$ atom at point $\boldr$, $\hat{H}_1 = \frac{-\hbar^2}{2m} \grad^2 $ and $\hat{H}_2 = \hat{H}_1 + \hbar \omega_2 $ are the single particle hamiltonians for state $|1\rangle$ and $|2\rangle$ atoms respectively,  $g_c$ is the vacuum Rabi-frequency for the optical cavity, and $\boldk_1$ and $\boldk_2$ are the wave vectors of the classical driving field and cavity mode respectively. Assuming that the state $|1\rangle$ atoms only occupy one motional state $\Psi_g(\boldr)$, the Hamiltonian describing the system is well-approximated by
\begin{eqnarray}
\mathcal{H} &=& \hbar \left(\omega_3 -\omega_2 -\Delta_2  \right) \chatd\chat \nonumber  \\
&+& \left(\frac{\hbar^2 (\boldk_1-\boldk_2)^2}{2m} + \hbar \omega_2 \right)\ahatd_2\ahat_2 \nonumber \\
&+& \hbar \left(\chi \ahat_1\ahatd_2\chatd e^{-i(\omega_3-\Delta_1)t} + {\rm h.c.} \right) \label{mainHam}
\end{eqnarray}
Here, $\chi = \frac{g_c \Omega_{13}}{\Delta_1}$, and 
\begin{subequations}
\begin{eqnarray}
\ahat_1 &=& \int \Psi^*_g(\boldr) \psihat_1(\boldr) \dr  \\ 
\ahat_2 &=& \int \Psi^*_g(\boldr) \psihat_2(\boldr) e^{i(\boldk_1-\boldk_2)\cdot \boldr} \dr \, ,
\end{eqnarray}
\end{subequations}
where $\Psi_g(\boldr)$ is the ground state single particle wavefunction of the BEC. 

To include the effects of coupling of the cavity mode to the environment, we use the standard input-output theory for optical cavities \cite{Walls:2008}. Furthermore, we also account for emission of photons into non-cavity modes. As each emission event corresponds to an atom receiving a momentum kick, which will transfer the atoms into a distinguishable momentum state, it is effectively a form of loss for condensate atoms. We account for this process phenomenlogically by adding a standard loss term to the condensate mode, proportional to the spontaneous emission rate \cite{Scully:1997}. Including these effects, the equations of motion are
\begin{subequations}
\begin{eqnarray}
i\dot{\ahat}_1 &=& \chi \ahat_2 \chat -i\frac{\gamma}{2} \ahat_1 + i\sqrt{\gamma} \ahat_{\mathrm{in}}\\
i\dot{\ahat}_2 &=& \chi \ahat_1 \chatd \\
i\dot{\chat} &=& \chi \ahat_1 \ahatd_2 -i\frac{\kappa}{2} \chat + i\sqrt{\kappa} \bhat_{\mathrm{in}} 
\end{eqnarray}
\label{EOMS}
\end{subequations}
where we have made the transformation 
\begin{subequations}
\begin{eqnarray}
\ahat_2 &\mapsto& \ahat_2 \exp(i\omega_a t) \\  
\chat &\mapsto& \chat \exp(i\omega_c t) , 
\end{eqnarray}
\label{change_of_variables}
\end{subequations}
where
\begin{subequations}
\begin{eqnarray}
\omega_a &=& \frac{\hbar (\boldk_1-\boldk_2)^2}{2m} + \omega_2 \\
\omega_c &=& \omega_3 -\omega_2 -\Delta_2
\end{eqnarray}
\end{subequations}
and assumed that the frequency of the classical driving field and the cavity mode were adjusted such that the system was on two-photon resonance:
\begin{equation}
\hbar(\Delta_2 - \Delta_1) = \frac{\hbar^2}{2m}\left( \boldk_1-\boldk_2 \right)^2 .
\end{equation}
The operators $\ahat_{\mathrm{in}}$ and $\bhat_{\mathrm{in}}$ satisfy
\begin{equation}
\left[\ahat_{\mathrm{in}}(t) \, , \ahatd_{\mathrm{in}}(t^\prime)\right] = \left[\bhat_{\mathrm{in}}(t) \, , \bhatd_{\mathrm{in}}(t^\prime)\right] = \delta(t-t^\prime) \, .
\end{equation}
Another quantity of interest is the output field $\bhat_{\mathrm{out}}(t)$, defined by
\begin{equation}
\bhat_{\mathrm{out}}(t) = \sqrt{\kappa}\chat(t) -\bhat_{\mathrm{in}}(t) \, .
\end{equation}
Here, $\gamma = \Gamma_{13} (\Omega_{13} /\Delta_1)^2$, where $\Gamma_{13}$ is the natural linewidth of the $|1\rangle \rightarrow |3\rangle$ transition, and $\kappa$ is the cavity linewidth. We begin with all of the atoms in mode $\ahat_1$, and then coherently transfer a small \emph{seed} of atoms to mode $\ahat_2$ via a coherent Raman transition. It is assumed that $\ahat_{\mathrm{in}}$ always operates on a vacuum state, as there is no physical process that will add atoms to our system, but $\bhat_{\mathrm{in}}$ can operate on any general state, to allow for the possibility of coherently driving the cavity field with an input laser.
 
\subsection{Measurement Stage}
After some time $t_1$, the pump laser is turned off such that the population dynamics terminates. We then use the atomic modes ($\ahat_1(t_1)$ and $\ahat_2(t_1)$) generated by this process  as the input to a standard Mach Zehnder (MZ) interferometer ($\frac{\pi}{2}-\pi-\frac{\pi}{2}$ sequence), where the pulses are implemented via coherent 2-photon Raman transitions with lasers of same frequencies and wave-vectors as used in the state preparation stage. Both optical modes are assumed to be sufficiently bright and coherent that it is sufficient to treat them classically. At $t=t_f$ (ie, after the MZ pulse sequence), we measure the number of particles in each mode, which is used to construct the \emph{signal} $\hat{S}$. 

The behaviour of an MZ interferometer is best understood by introducing the pseudospin operators 
\begin{subequations}
\begin{eqnarray}
\jhat_x &=& \frac{1}{2}\left(\ahat_1^\dag\ahat_2 + \ahat_1\ahat_2^\dag \right) \\
\jhat_y &=& \frac{i}{2}\left(\ahat_1\ahat_2^\dag - \ahat_1^\dag\ahat_2\right) \\
\jhat_z &=& \frac{1}{2}\left(\ahat_1^\dag\ahat_1 - \ahat_2^\dag\ahat_2\right) \notag \\
&=& \frac{1}{2}\left(\hat{N}_1 - \hat{N}_2\right)
\end{eqnarray}
\end{subequations}
where $\hat{N}_1$($\hat{N}_2$) is the population in mode $\ahat_1$($\ahat_2$). The MZ interferometer performs the unitary transformation $\hat{U}_\phi  = \exp \left( -i \phi \jhat_y\right)$, where $\phi$ is the accumulated phase difference between the modes, such that 
\begin{eqnarray}
\hat{J}_z(t_f) &=& \hat{U}^\dag_\phi \jhat_z(t_1) \hat{U}_\phi \nonumber \\
&=& \hat{J}_z(t_1) \cos \phi - \hat{J}_x(t_1) \sin \phi \, .
\end{eqnarray}
For a given quantum state $|\Psi (t_1)\rangle$ input to the device, the smallest phase shift $\phi$ resolvable by the device is given by the Quantum Cramer-Rao Bound (QCRB) \cite{Braunstein:1994, Paris:2009, Demkowicz-Dobrzanski:2014, Toth:2014}
\begin{equation}
\Delta \phi_\text{QCRB} = \frac{1}{\sqrt{\mathcal{F}_Q}} \,
\end{equation}
where $\mathcal{F}_Q$ is the quantum fisher information, which for pure states can be calculated by
\begin{equation}
\mathcal{F}_Q = 4\left(\langle \partial_\phi \Psi_\phi |  \partial_\phi \Psi_\phi \rangle- \lvert \langle \partial_\phi \Psi_\phi |  \Psi_\phi \rangle \rvert^2  \right ) \, ,
\end{equation}
where
\begin{equation}
|  \partial_\phi \Psi_\phi \rangle = \frac{d}{d\phi} \left(\hat{U}_\phi |\Psi(t_1)\rangle\right) \, .
\end{equation}
Using $\hat{U}_\phi =  \exp \left( -i \phi \jhat_y\right)$ gives
\begin{equation}
\mathcal{F}_Q = 4 V(\jhat_y) \, ,\label{fisher_def}
\end{equation}
where the variance is calculated with respect to the input state $|\Psi(t_1)\rangle$. 

However, for a given choice of measurement signal, $\hat{S}$, the phase sensitivity of the device is
\begin{equation}
\Delta \phi = \frac{\xi_S}{\sqrt{N_t}} \label{delta_phi_def}
\end{equation}
where
\begin{equation}
\xi_S = \sqrt{N_t}\frac{\sqrt{V(S)}}{|\partial_\phi \langle \hat{S}\rangle|} \label{xi_def}
\end{equation}
and $V(S) = \langle \hat{S}^2\rangle - \langle \hat{S}\rangle^2$ is the variance of the signal. We refer to $\xi_S$ as the quantum-enhancement parameter, as it quantifies the phase sensitivity relative to the SQL; $\xi_S < 1$ indicates sensitivity better than the SQL. A measurement scheme is optimum when it saturates the QCRB, ie $\Delta \phi = \Delta \phi_\text{QCRB}$, or equivalently, $\xi_S = \xi_F$, where 
\begin{equation}
\xi_F = \frac{\sqrt{N}_t}{\sqrt{\mathcal{F}_Q}} \, .
\end{equation}

Taking $\hat{S} = \jhat_z(t_f)$, at the most sensitive phase for our choice of input state, $\phi = \frac{\pi}{2}$, we find
\begin{equation}
\xi_S = \frac{\sqrt{N_t} \sqrt{V(\jhat_x(t_1))}}{| \langle \jhat_z(t_1)\rangle|} \label{delta_phi_J}
\end{equation}
However, as in \cite{Haine:2013, Szigeti:2014b, Tonekaboni:2015, Haine:2015, Haine:2015b}, we can gain an enhancement by using \emph{information recycling}: if the atomic degrees of freedom are correlated with the optical field, it may be possible to gain an enhancement by incorporating measurements of the optical field into our signal. Specifically, we choose 
\begin{equation}
\hat{S} = \jhat_z(t_f) + \hat{S}_b \, ,
\end{equation}
where $\hat{S}_b$ is some observable that involves only the photonic degrees of freedom of the system. Noting that measurements of $\hat{S}_b$ are independent of $\phi$, at $\phi = \pi/2$ we find 
\begin{equation}
\xi_S = \frac{\sqrt{N_t}\sqrt{V\left(-\jhat_x(t_1) + \hat{S}_b(t_1)\right)}}{|\langle \jhat_z(t_1) \rangle|} \, . \label{xi_info}
\end{equation}
Obviously, $V(-\jhat_x(t_1) + \hat{S}_b(t_1)) \geq V(\jhat_x(t_1))$ when the atomic and photonic systems are seperable. However,  when there is atom-light entanglement in the system, there may be some photonic operator $\hat{S}_b$ such that $V(S)$ is reduced and the sensitivity is increased over purely atomic measurements. It is interesting to note that when there is atom-light entanglement, ignoring information in the photonic degrees of freedom is equivalent to tracing over these degrees of freedom, and the state of the system is no longer pure, and the Quantum Fisher information cannot be calculated via \eq{fisher_def}. However, if this information is incorporated into the signal, then the system can be treated as pure, and \eq{fisher_def} remains appropriate \cite{Haine:2015b}. 

In the next section, we will model the dynamics of the system and investigate how the choice of $\hat{S}_b$ effects $\xi_S$. 

\section{Cavity Dynamics }
\subsection{Perfect Cavity}
We begin by analysing the simplified case of a perfect cavity, and also neglect the effects of spontaneous emission by setting $\kappa\rightarrow 0$, $\gamma \rightarrow 0$ in Eqs.~\ref{EOMS}. In \cite{Haine:2013}, as we did not include an optical cavity, a ``seed" in mode $\ahat_2$ was required to stimulate transitions into this mode. A minimum value of the seed was required in order for the stimulated processes to dominate the spontaneous emission. In a perfect cavity, however, there is nothing to prevent us setting the seed size to zero, assuming that the effective cavity coupling rate $\chi \sqrt{N_t}$, where $N_t$ is the total number of atoms, is large compared to the spontaneous emission rate $\gamma$. The use of a cavity also gives us the freedom to use a seed in mode $\chat$. 

We can gain an understanding of how the correlations can enhance the interferometry, and what is a useful choice for $\hat{S}_b$ with a simplified model. Treating mode $\ahat_1$ as a large undepletable reservoir, we can make the undepleted pump approximation $\ahat_1 \rightarrow \sqrt{N_{a_1}}$, yielding the simplified equations of motion
\begin{subequations}
\begin{eqnarray}
i\dot{\ahat}_2 &=& \chi \sqrt{N_t} \chatd \\
i\dot{\chat} &=& \chi \sqrt{N_t} \ahatd_2 \, ,
\end{eqnarray}
\end{subequations}
which has solution
\begin{subequations}
\begin{eqnarray}
\ahat_2(t) &=&  \ahat_2(0)\cosh r - i \chatd(0) \sinh r \\
\chat(t) &=&  \chat(0)\cosh r - i \ahatd_2(0) \sinh r \, ,
\end{eqnarray}
\label{ahat_anal}
\end{subequations}
where $r = \sqrt{N}_t\chi t$. It is well known from quantum optics \cite{Scully:1997,Walls:2008,Bachor:2004} that such dynamics leads to correlations between the amplitude and phase quadratures of modes $\ahat_2$ and $\chat$. Specifically, when $|\Psi(t_0)\rangle = |0\rangle$, 
\begin{equation}
V\left( \hat{X}_{a_2} - \hat{Y}_{c}\right) = 2 e^{-2r} \, 
\end{equation}
where 
\begin{subequations}
\begin{eqnarray}
\hat{X}_{a_2} &=& \ahat_2 + \ahatd_2 \\
\hat{Y}_{c} &=& i(\chat - \chatd) \, .
\end{eqnarray}
\end{subequations}
In order to minimize \eq{xi_info}, we notice that in the undepleted pump limit, $\jhat_x \approx \sqrt{N_{a_1}}\hat{X}_{a_2}$, and therefore setting $\hat{S}_b = \sqrt{N_{a_1}}\hat{Y}_c$ gives
\begin{equation}
\xi_S \approx \frac{\sqrt{2 e^{-2r} N_t\left(N_t -\sinh^2 r \right)} }{\lvert N_t - 2\sinh^2 r \rvert} \sim \sqrt{2} e^{-r} \, ,
\end{equation}
where we have enforced conservation of the total number of atoms via $N_{a_1} = N_t - \langle \ahatd_2\ahat_2\rangle$. 
When $|\Psi(t_0)\rangle$ has a non-zero component in either $\ahat_2$ or $\chat$, the expression is more complicated, and depends on the relative phases of these two coherent seeds. 

To investigate the effect of depletion from $\ahat_1$, we use proceed by using the truncated Wigner (TW) approximation \cite{Walls:2008, Gardiner:2004b}. Following standard methods \cite{Steel:1998, Blakie:2008}, the Heisenberg equations (Eqs.~\ref{EOMS}) can be converted into Fokker-Plank equations (FPEs) by using the correspondences between the quantum operators and the Wigner function. By truncating third- and higher-order terms, the FPEs can be mapped onto a set of stochastic partial differential equations for complex valued variables $\alpha_1$, $\alpha_2$, $\beta_{\mathrm{in}}$, and $\mathcal{C}$, which we solve numerically. The stochastic differential equations describing the evolution of the system are
\begin{subequations}
\begin{eqnarray}
i\dot{\alpha}_1 &=& \chi \alpha_2 \mathcal{C} -i\frac{\gamma}{2} \alpha_1 + i\sqrt{\gamma} \alpha_{\mathrm{in}}\\
i\dot{\alpha}_2 &=& \chi \alpha_1 \mathcal{C}^* \\
i\dot{\mathcal{C}} &=& \chi \alpha_1 \alpha_2^* -i\frac{\kappa}{2} \mathcal{C} + i\sqrt{\kappa} \beta_{\mathrm{in}} 
\end{eqnarray}
\label{wigEOMS}
\end{subequations}
where we  have made the operator correspondences $\ahat_{1(2)} \rightarrow \alpha_{1(2)}$, $\bhat_{\mathrm{in}}\rightarrow \beta_{\mathrm{in}}$, and $\chat \rightarrow \mathcal{C}$. By averaging over many trajectories with initial conditions sampled from the appropriate Wigner function, expectation values of quantities corresponding to operators in the full quantum theory can be obtained \cite{Steel:1998, Olsen:2009}. Specifically,
\begin{equation}
\langle \{ f\left( \ahatd_1, \ahat_1, \ahatd_2, \ahat_2, \chatd, \chat \right) \}_\mathrm{sym}\rangle = \overline{f\left( \alpha^*_1, \alpha_1, \alpha^*_2, \alpha_2, \mathcal{C}^*, \mathcal{C} \right) } \, ,
\end{equation}
where ``sym" denotes symmeteric ordering, and the overline denotes the mean over many stochastic trajectories. 
 We typically assume that the initial state of each mode of the field is a Glauber coherent state:
\begin{equation}
|\Psi(t_0)\rangle = \mathcal{D}_{a_1}\mathcal{D}_{a_2}\mathcal{D}_{c}|0\rangle \, , \label{psi_init}
\end{equation}
where
\begin{subequations}
\begin{eqnarray}
\mathcal{D}_{a_1} &=& \exp\left(\alpha_{10}\ahatd_1 - \alpha^*_{10}\ahat_1 \right) \\
\mathcal{D}_{a_2} &=& \exp\left(\alpha_{20}\ahatd_2 - \alpha^*_{20}\ahat_2 \right) \\
\mathcal{D}_{c} &=& \exp\left(\mathcal{C}_0\chatd - \mathcal{C}_0^*\chat \right) \, .
\end{eqnarray}
\end{subequations}
One subtlety of this choice of initial state is that massive particles obey super-selection rules, which strictly forbid the possibility of coherent superpositions of different numbers of particles, such as are present in the Glauber coherent state. However, it was shown in \cite{Haine:2009} that a \emph{mixture} of coherent states with randomized phase (which corresponds to a Possonian mixture of number states) behaves identically to a pure coherent state for the purposes of atom interferometry. This is true even for the choice of a non-zero seed in mode $\ahat_2$, as long as it was created by coherently transferring atoms from mode $\ahat_1$. In this case, while the phase of both $\alpha_{10}$ and $\alpha_{20}$ are random, the \emph{relative} phase is not.

We also allow for the possibility of driving the cavity with a either pulsed or continuous coherent light. Specifically, we assume that the state of the EM field outside the cavity is
\begin{equation}
|\Psi_{\mathrm{in}}(t)\rangle = \exp\left(\beta_0(t)\bhatd - \beta_0^*(t)\bhat_1 \right)|0\rangle \,
\end{equation} 
where the time dependence in $\beta_0$ allows for the possibility of temporal dynamics in the amplitude of the coherent state. This leads to the initial condition for the stochastic differential equations of
\begin{subequations}
\begin{eqnarray}
\alpha_1(0) &=& \alpha_{10} + \eta_1 \\
\alpha_2(0) &=& \alpha_{20} + \eta_2 \\
\mathcal{C}(0) &=& \mathcal{C}_{0} + \eta_3 \\
\beta_{\mathrm{in}}(t) &=& \beta_0(t) + w_{\beta}(t) \\
\alpha_{\mathrm{in}}(t) &=& w_{\alpha}(t)
\end{eqnarray}
\end{subequations}
where $\eta_j$ is complex gaussian noise satisfying $\overline{\eta_j} = 0$, and $\overline{\eta^*_i\eta_j} = \frac{1}{2}\delta_{ij}$, and $w_\nu(t)$ is a complex Wiener noise satisfying $\overline{w_\nu(t)} = 0$, and $\overline{w_\mu^*(t)w_\nu(t^\prime)} = \frac{1}{2}\delta_{\mu\nu}\delta(t-t^\prime)$

\begin{figure}
\includegraphics[width=1\columnwidth]{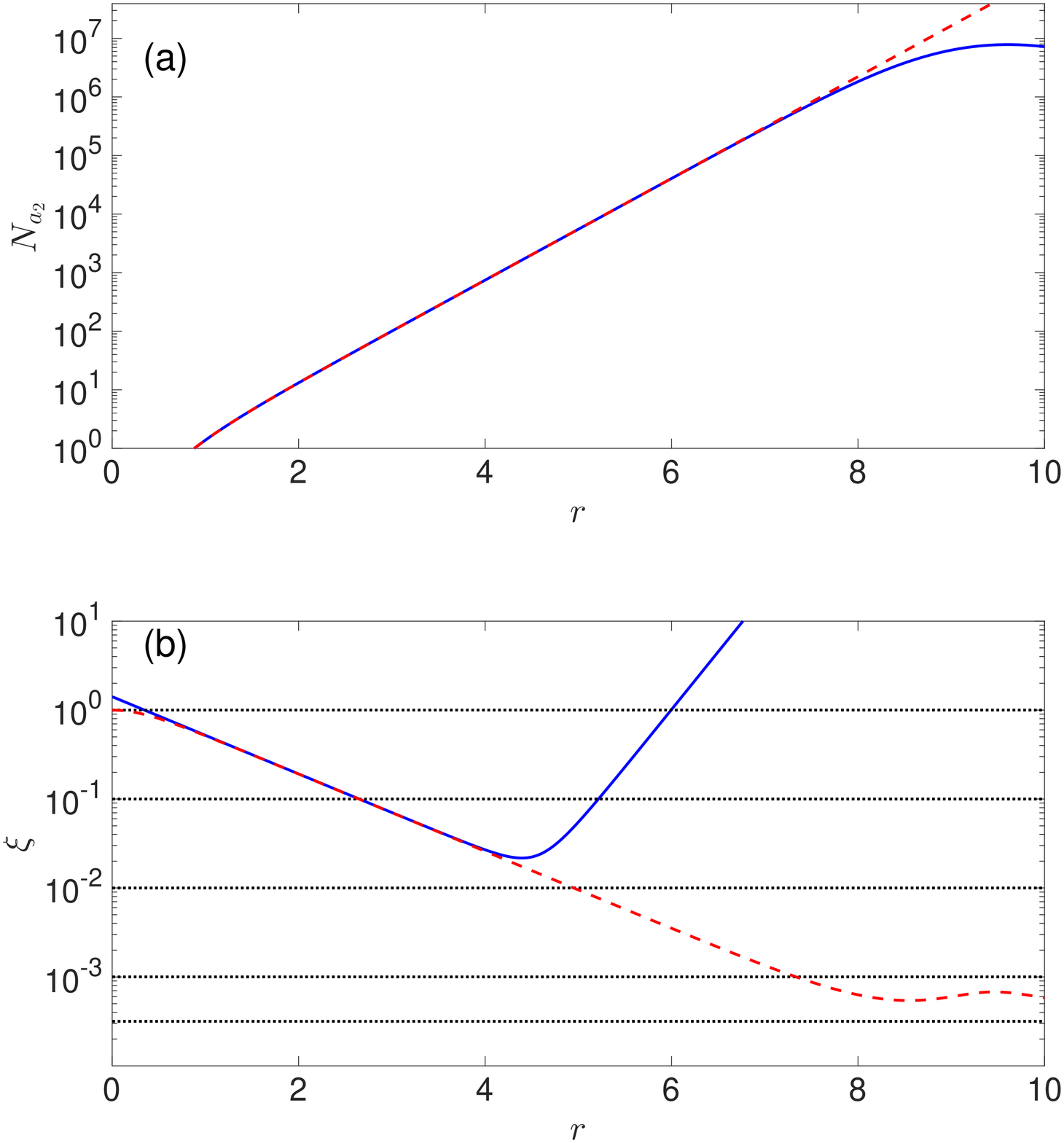}
\caption{(Color online)  (a): $N_{a_2} = \langle \ahatd_2\ahat_2\rangle$ vs. $r = \chi \sqrt{N_t} t$ calculated via equations (\ref{wigEOMS}), (blue solid line) and \eq{ahat_anal} (red dashed line), for an initial state such that $\langle \ahatd_2\ahat_2\rangle = \langle \chatd\chat\rangle = 0$. (b) $\xi_S$ (blue solid line) and $\xi_F$ (red dashed line). The black dotted lines are simply to guide the eye, with the lowest line indicating $\Delta \phi = \frac{1}{N_t}$. Parameters: $\alpha_{10} = \sqrt{10^7}$, $\alpha_{20} = \mathcal{C}_0 = 0$. 
}
\label{pops_noseed}
\end{figure}

We begin by investigating the perfect cavity case ($\gamma \rightarrow 0$, $\kappa \rightarrow 0$), with no seed in either $\ahat_2$ or $\chat$ ($\alpha_{20} = 0$, $\mathcal{C}_0=0$). We solved \eqs{wigEOMS} numerically. Fig (\ref{pops_noseed}) (a) shows the population in $\ahat_2$ calculated from \eqs{wigEOMS} compared to the analytic solution from \eqs{ahat_anal}. For short times, there is excellent agreement in the population between the undepleted pump approximation, and the TW simulation. Fig (\ref{pops_noseed}) (b) shows $\xi_S$ and $\xi_F$ as a function of preparation time using 
\begin{equation}
\hat{S} = \jhat_z(t_f) + \frac{1}{2} \sqrt{\langle \ahatd_1 \ahat_1\rangle} \hat{Y}_c \label{sig_cav}
\end{equation} 
as the signal.  For $r> \log \sqrt{2} \sim 0.35$, $\xi_S < 1$, indicating sub-SQL sensitivities. We have also calculated $\xi_F$. As $r$ becomes greater than  $\sim \log \sqrt{2}$, $\xi_S \approx \xi_F$, indicating that our choice of signal is close to optimum for the choice of quantum state. However, as $r$ becomes larger than $\sim 4$, $\xi_S$ and $\xi_F$ begin to diverge. The optimum sensitivity is approximately $\xi_S \approx 0.021$, while the optimum allowed by the QCRB is $\xi_F \approx 5.4 \times 10^{-4}$, which leads to a sensitivity of $\Delta \phi_\text{QCRB} \approx 1.7 /N_t$, which is very close to the Heisenberg limit \cite{Holland:1993, Giovannetti:2006}. This indicates that for large values of $r$ our signal is not optimal, and there is some better measurement or method of processing the information. It has been shown in \cite{Haine:2015b} that the signal
\begin{equation}
\hat{S} = \left(\hat{J}_z(t_f) + \chatd\chat - \frac{1}{2}(\ahatd_1\ahat_1 + \ahatd_2\ahat_2)\right)^2 \label{kralken_sig}
\end{equation}
saturates the QCRB. However, for moderate values of $r$, \eq{sig_cav} is \emph{almost} optimal, and is sufficient to provide significant quantum enhancement. Furthermore, it is considerably simpler to work with in practice, as at the optimum phase shift, $\partial_\phi \langle \hat{S} \rangle$ is large, unlike \eq{kralken_sig}, for which $\partial_\phi \langle \hat{S} \rangle \rightarrow 0$ at the optimum phase. Additionally, homodyne detection of quantum optical correlations is reasonably routine, while high-efficiency photon-counting with single photon resolution is somewhat challenging \cite{Bachor:2004}. 

\begin{figure}
\includegraphics[width=1\columnwidth]{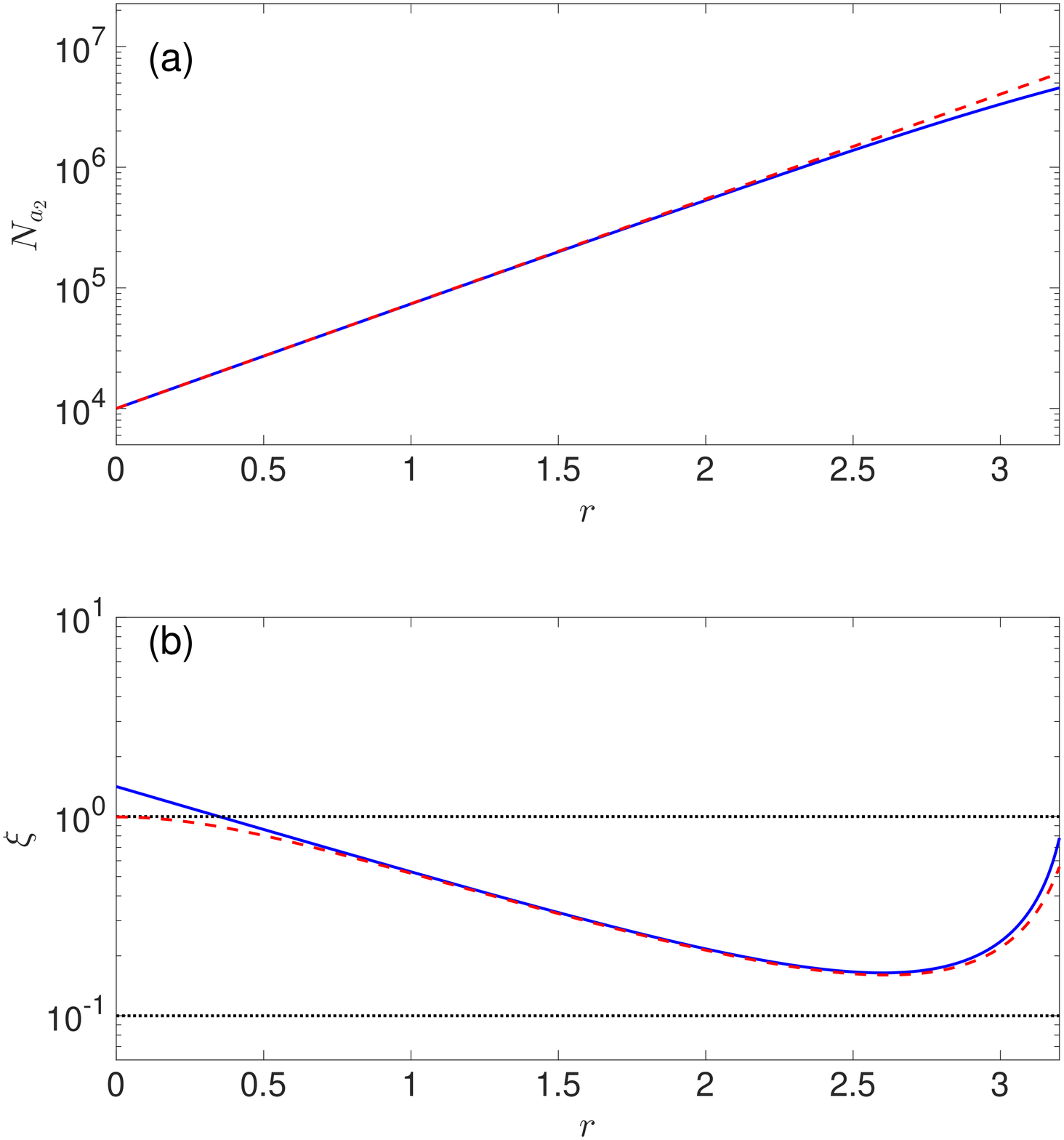}
\caption{(Color online) (a): $N_{a_2} = \langle \ahatd_2\ahat_2\rangle$ vs. $r = \chi \sqrt{N_t} t$ calculated via equations (\ref{wigEOMS}), (blue solid line) and \eq{ahat_anal} (red dashed line), for the initial state \eq{psi_init}, with $\alpha_{20} = -i\sqrt{N_\text{seed}}$, $\mathcal{C}_0 = \sqrt{N_\text{seed}}$, and $\alpha_{10} = \sqrt{N_t - N_\text{seed}}$. (b) $\xi_S$ (blue solid line) and $\xi_F$ (red dashed line). Parameters: $N_t = 10^7$, $N_{\mathrm{seed}} = 10^4$. 
}
\label{pops_symseed}
\end{figure}

There may be some situations in which the use of a coherent seed in either mode $\ahat_2$ or $\chat$ is desirable, such as when the effective cavity coupling constant is insufficient to overwhelm the spontaneous emission rate. 
Fig.~\ref{pops_symseed} shows the population in $\ahat_2$ and $\xi_S$ when the initial population in $\ahat_2$ and $\chat$ were chosen such that $\langle \ahatd_2 \ahat_2\rangle = \langle \chatd_2 \chat_2\rangle = N_\text{seed}$. We can see that the use of the seed causes the population to grow much more rapidly. However, the use of a seed inhibits the degree of quantum enhancement, with the $\text{min}(\xi_F) \approx \text{min}(\xi_S) \approx 0.16$, indicating that with the use of a large coherent seed, \eq{sig_cav} is a good approximation to the optimal signal. The degradation in quantum enhancement is due to the initial seed acting as a source of uncorrelated particles, which reduces the degree of correlations between mode $\ahat_2$ and $\chat$. 

\begin{figure}
\includegraphics[width=1\columnwidth]{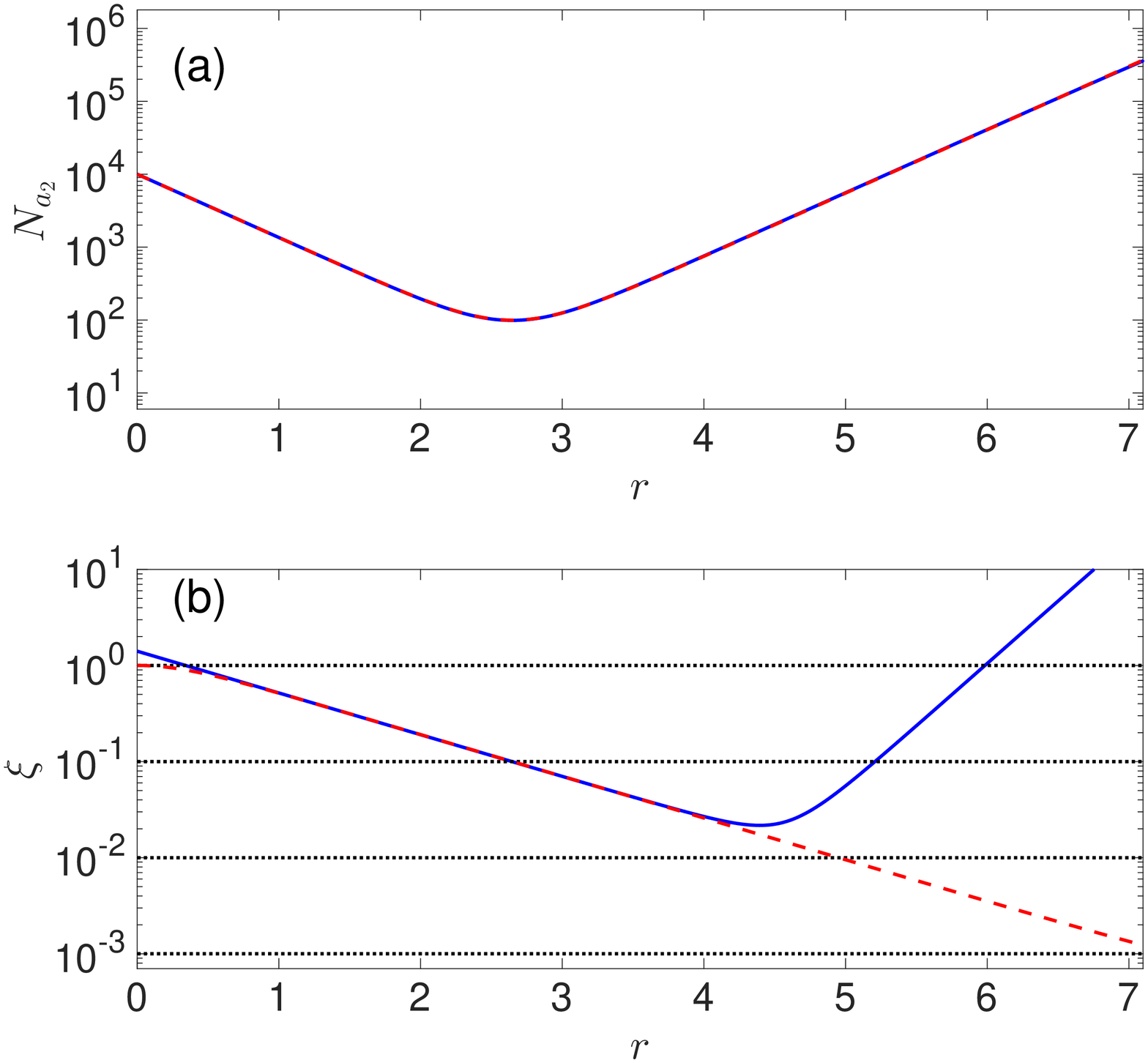}
\caption{(Color online) (a): $N_{a_2} = \langle \ahatd_2\ahat_2\rangle$ vs. $r = \chi \sqrt{N_t} t$ calculated via equations (\ref{wigEOMS}), (blue solid line) and \eq{ahat_anal} (red dashed line), for the initial state \eq{psi_init}, with $\alpha_{20} = -i\sqrt{N_\text{seed}}$, $\mathcal{C}_0 = -\sqrt{N_\text{seed}}$, and $\alpha_{10} = \sqrt{N_t - N_\text{seed}}$. (b) $\xi_S$ (blue solid line) and $\xi_F$ (red dashed line). Parameters: $N_t=10^7$, $N_{\mathrm{seed}} = 10^4$. 
}
\label{pops_asymseed}
\end{figure}

The choice of the \emph{phase} of the initial seed in each mode can also affect the dynamics. In the previous example we chose $\alpha_{20} = -i\sqrt{N_\text{seed}}$, and $\mathcal{C}_{0} = \sqrt{N_\text{seed}}$. The motivation for this choice was that a seed in $\ahat_2$ resulting from coherently transferring atoms from $\ahat_1$ via a two-photon Raman transition implemented by two coherent lasers with no relative phase difference between them results in a relative phase between the two atomic modes of $-\pi/2$. The phase of the seed in $\chat$ is arbitrary, but a phase of zero indicates that this field is in phase with the laser driving the $|1\rangle\rightarrow |3\rangle$ transition. However, applying a $\pi$ phase shift to the seed in $\chat$ results in significantly different dynamics. Fig.~\ref{pops_asymseed}(a) shows the population dynamics in this case. In this situation, the atom-light coupling initially results in \emph{de-amplification} of the initial seed. After this the population is then amplified, and grows at a similar rate to the case with no seed. Both $\xi_S$ and $\xi_F$ mimic the seedless case. It is tempting to think that in this regime we have both the benefits of a large seed (reduction in spontaneous emission) and a large QFI. However, as the population is reduced before it grows, it takes approximately the same amount of time to reach the optimum, so will suffer approximately the same degree of spontaneous emission. 

\begin{figure}
\includegraphics[width=1\columnwidth]{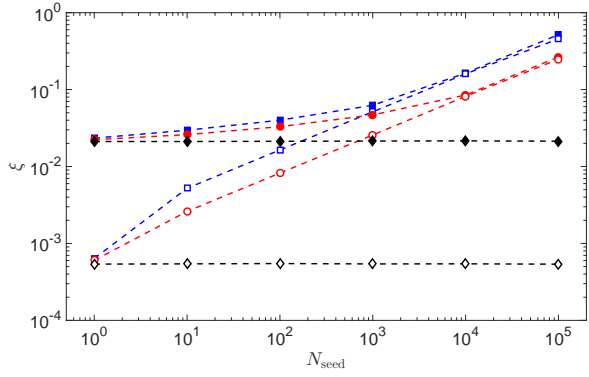}
\caption{(Color online) ).  Minimum obtainable $\xi_F$ (hollow shapes) and $\xi_S$ (solid shapes) vs. $N_\text{seed}$ for $\alpha_{20} = -i\sqrt{N_\text{seed}}$, $\mathcal{C} = \sqrt{N_\text{seed}}$ (blue squares), $\alpha_{20} = -i\sqrt{N_\text{seed}}$, $\mathcal{C} = -\sqrt{N_\text{seed}}$ (black diamonds), and $\alpha_{20} = -i\sqrt{N_\text{seed}}$, $\mathcal{C} = 0$ (red circles). 
}
\label{xi_vs_seed}
\end{figure}

Fig.~\ref{xi_vs_seed} shows the minimum obtainable $\xi_S$ and $\xi_F$ vs. $N_\text{seed}$ for different relative phases of $\alpha_{20}$ and $\mathcal{C}_0$. There is a general trend that as the seed is larger, the minimum of $\xi_F$ increases, and $\xi_S$ approaches $\xi_F$.  The case when  \{$|\alpha_{20}|^2 = 0$, $|\mathcal{C}_0|^2 =N_{\rm seed}$\} provides identical sensitivity to the \{$|\alpha_{20}|^2 = N_{\rm seed}$, $|\mathcal{C}_0|^2 =0$\} case, so has been omitted for visual clarity. 

Obviously, a perfect cavity is unrealistic, and it is unfeasible to directly make measurements on the cavity mode. To model a realistic system we must take into account coupling of the cavity mode to the freely propagating continuum.  
In the next section we will investigate the behaviour of a realistic cavity, including coupling between freely propagating modes and the cavity mode, and the effects of spontaneous emission. 

\subsection{Realistic Cavity}
\begin{figure}
\includegraphics[width=1\columnwidth]{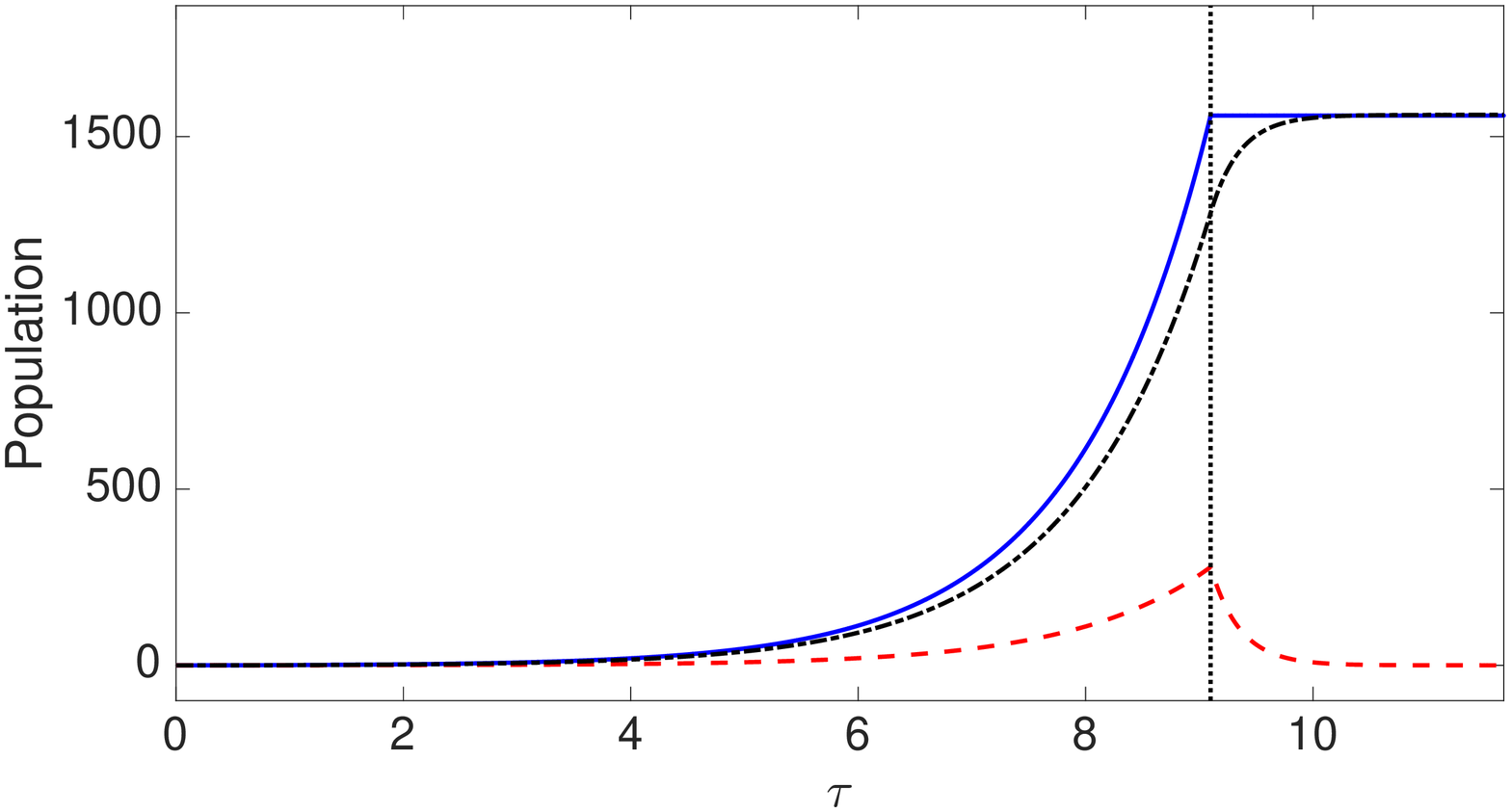}
\caption{(Color online).  $N_{a_2}$ (blue solid line), $N_c$ (red dashed line), 
and $N_b$ (black dot-dashed line) vs. $\tau = \chi \sqrt{N_t} t$. The atom-light coupling, $\chi$ was set to zero at $\tau = 9.1$, which was chosen as it roughly corresponds to the time when $N_{a_2}$ is equal to the number of transferred atoms when $\xi_S$ is minimum in Fig.~\ref{pops_noseed}. Parameters: $\alpha_{10} = \sqrt{10^7}$, $\alpha_{20} = \mathcal{C}_0 = 0$, $\beta_0=0$, $\kappa = 13.0 \times 10^6$ rad s$^{-1}$, $\chi = 1.06 \times 10^3$ rad s$^{-1}$, and $\gamma = 600$ rad s$^{-1}$.
}
\label{pops1}
\end{figure}

The previous section ignored the coupling between the cavity and the environment, as well as the effect of spontaneous emission. In this section we investigate the behaviour of the system in the presence of these effects, and how information recycling can be implemented. We begin by investigating the behaviour of the system in the absence of a seed. We solved \eqs{wigEOMS} numerically, using the relevant cavity QED parameters $\{g_c, \kappa, \Gamma_{13}\} = \{0.106, 13.0, 6.0\} \times 10^6$ rad s$^{-1}$. Here, $\Gamma_{13}$ corresponds to the spontaneous emission rate of the $D_2$ line of $^{87}$Rb, and $g_c$ and $\kappa$ correspond to a cavity that is not in the strongly coupled cavity QED regime where $g_c > (\kappa, \Gamma_{13})$. The values of  $g_c$ and $\kappa$ that we have used are a factor of $100$ less and $10$ greater than was reported in \cite{Brennecke:2007}, respectively. Our motivation for such a choice was that while these values would be challenging to achieve, they are not `state of the art' and therefore not an unreasonable modification to an existing atom-interferometry setup. Setting $\Omega_{13} / \Delta_1 = 10^{-2}$ gives $\chi = 1.06 \times 10^3$ rad s$^{-1}$, and $\gamma = 600$ rad s$^{-1}$. Fig.~\ref{pops1} shows the populations $N_{a_2} = \langle \ahatd_2 \ahat_2\rangle$, $N_c = \langle \chatd \chat\rangle$ and $N_b =  \int_0^t \langle \bhatd_{\mathrm{out}}(t^\prime)\bhat_{\mathrm{out}}(t^\prime)\rangle dt^\prime$, calculated by numerically solving \eqs{wigEOMS}. For these parameters, the atomic field grows significantly faster than the cavity field, as photons leak out of the cavity. 

In a real experiment we do not have access to the cavity mode $\chat$, and in order to implement information recycling, we must rely on the measurements of the output field $\bhat_{\mathrm{out}}$.
 In analogy with the perfect cavity case, information recycling is implemented by using the signal
\begin{equation}
\hat{S} = \jhat_z(t_f) + \frac{1}{2} \sqrt{\langle \ahatd_1 \ahat_1\rangle} \hat{Y}_b \label{sig_cavb}
\end{equation}
where $\hat{Y}_b = i(\bhat_0 -\bhatd_0)$ is the phase quadrature of a specific mode of the output field defined by
\begin{equation}
\bhat_0 = \int_0^{T} u^*_{\mathrm{LO}}(t) \bhat_{\mathrm{out}}(t) dt \, .
\end{equation}
Physically, the mode function $u_{\mathrm{LO}}(t)$ corresponds to the temporal mode shape of the `local oscillator', or bright coherent state used in the homodyne detection of the output field \cite{Bachor:2004}. The choice of this function can significantly affect the correlations between $\bhat_0$ and $\ahat_2$. In order for $\hat{Y}_b$ to satisfy quadrature-like uncertainty relations, we require
\begin{equation}
\int_0^T \lvert u_{\mathrm{LO}}(t) \rvert ^2 dt =1 \, ,
\end{equation}
which implies $[ \bhat_0 \, , \, \bhatd_0 ] = 1$. For the effective implementation of information recycling, we require efficient transfer of information from the cavity mode to the output mode $\bhat_{\mathrm{out}}$. To demonstrate how the choice of $u_{\mathrm{LO}}(t)$ will influence this, we first consider the case with no atom-light coupling ($\chi=0$), where at $t=0$, the optical cavity contains some arbitrary quantum state $|\Psi_c\rangle$, while the state of the field outside the cavity is vacuum: $|\Psi(0)\rangle = |\Psi_c\rangle\otimes |0\rangle$. Using the method presented in \cite{Haine:2005b, Haine:2005}, we can express the solution to \eqs{EOMS} as
\begin{eqnarray}
\chat(t) &=& f(t) \chat(0) + \hat{v}_c(t) \\
\bhat_{\mathrm{out}}(t) &=& \sqrt{\kappa}f(t) \chat(0) + \hat{v}_b(t) \, ,
\end{eqnarray}
where $\hat{v}_c(t)$ and $\hat{v}_b(t)$ are chosen to preserve the commutation relations of $\chat(t)$ and $\bhat_{\rm out}(t)$, and have the property that $\hat{v}_{b(c)}(t)|\Psi(0)\rangle = 0$. Using this result, we find
\begin{eqnarray}
V(\hat{Y}_b(t)) &=& 2 |\eta|^2 \langle \chatd(0) \chat(0)\rangle +1 - (\eta^2 \langle \chat(0)^2 \rangle +h.c.) \nonumber \\
&+& (\eta \langle \chat(0)\rangle - h.c.)^2 \, ,
\end{eqnarray}
where 
\begin{equation}
\eta = \int_0^t u_{\mathrm{out}}^*(t^\prime) \sqrt{\kappa} f(t^\prime)\, dt^\prime \, .
\end{equation}
By choosing $u_{\mathrm{out}}^*(t)$ such that $\eta$ is real, this simplifies to
\begin{equation}
V(\hat{Y}_b(t)) = \eta^2 V(\hat{Y}_c(0)) + (1- \eta^2)
\end{equation}
We can see that as $\eta \rightarrow 1$, $V(\hat{Y}_b(t)) \rightarrow V(\hat{Y}_c(t))$, indicating that correlations initially contained in $\hat{Y}_c$ are efficiently transferred to $\hat{Y}_b$. When $\eta$ is complex, this corresponds to a rotation of the quadrature of the cavity mode. Clearly, the normalised function $u_{\mathrm{LO}}(t)$ that maximises $\eta$ is $u_{\mathrm{LO}}(t) \propto f(t) \propto \langle \chat(t)\rangle \propto \langle \bhat_{\mathrm{out}}(t)\rangle$. 

\begin{figure}
\includegraphics[width=1\columnwidth]{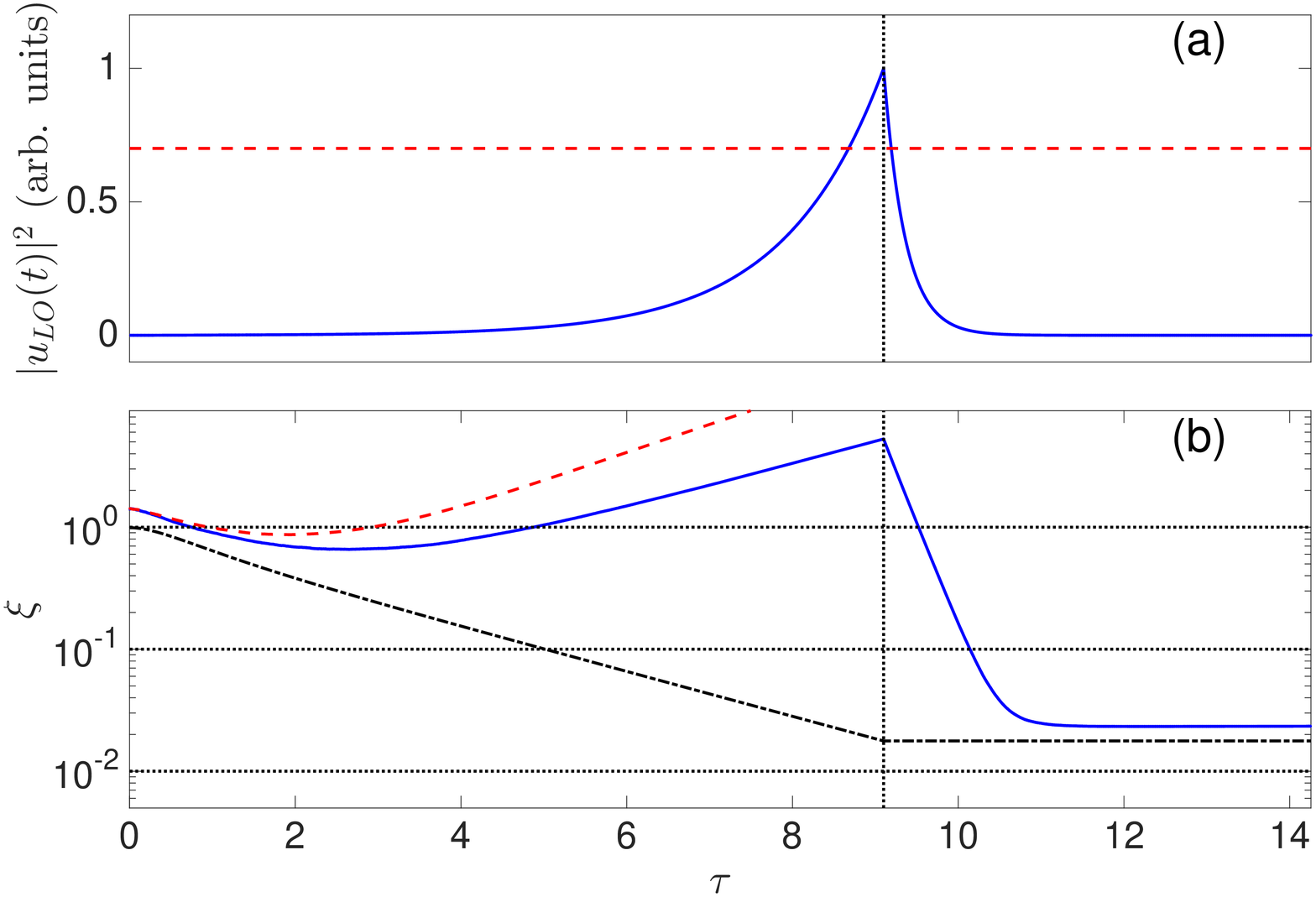}
\caption{(Color online).  Dynamics of a realistic cavity in the absence of a seed. The atom-light coupling was switched off at $\tau = 9.1$, as indicated by the vertical black dotted line in both figures. (a) The temporal shaping of $|u_{\mathrm{LO}}(t)|^2$ (blue solid line) used to calculate $\xi_S$. For comparison, we have also shown $u_{\mathrm{LO}}(t) \propto 1$ (red dashed line). 
(b) $\xi_S$ using \eq{u_opt1} (blue line), and $u_{\mathrm{LO}}(t) \propto 1$ (red dashed line). For each point on the curve, the normalization condition $\int_0^t |u_{\mathrm{LO}}(t^\prime)|^2 dt^\prime = 1$ was enforced. For comparison, we have included $\xi_F = \sqrt{N_t}/\sqrt{4 V(J_y)}$ (black dot-dashed line). Parameters: $\alpha_{10} = \sqrt{10^7}$, $\alpha_{20} = \mathcal{C}_0 = 0$, $\beta_0=0$, $\kappa = 13.0 \times 10^6$ rad s$^{-1}$, $\chi = 1.06 \times 10^3$ rad s$^{-1}$, and $\gamma = 600$ rad s$^{-1}$.
}
\label{delta_phi_realcav1}
\end{figure}

In the presence of atom-light coupling, \eqs{EOMS} are no longer linear and we can no longer make a simple linear ansatz for the solution. However, we can use our insight from the previous example to postulate that a good choice for $u_{\mathrm{LO}}(t)$ is proportional to the field of the cavity mode, or since $\langle \chat(t)\rangle = 0$, 
\begin{equation}
u_{\mathrm{LO}}(t) \propto \sqrt{\langle \chatd(t)\chat(t) \rangle} \propto \sqrt{\langle \bhatd_{\mathrm{out}}(t)\bhat_{\mathrm{out}}(t) \rangle}  \, . \label{u_opt1}
\end{equation}
Physically, this would require matching the temporal shape of the local oscillator to the expectation value of the intensity of the output field. It is assumed that the carrier frequency of the local oscillator $\omega_{\mathrm{LO}}$ is the same as the cavity mode $\omega_c$, which has automatically been accounted for in our change of variables (Eqs.~\ref{change_of_variables}). Fig.~\ref{delta_phi_realcav1} shows $\xi_S$, from \eq{sig_cavb}, using \eq{u_opt1}, for the same parameters as used in Fig.~\ref{pops1}. After most of the light leaking out of the cavity has been collected, $\xi_S \approx 0.023$. This is close to the limit set by the QCRB: $\xi_F = \sqrt{N_t}/\sqrt{4 V(J_y)} \approx 0.018$. We note that due to the presence of spontaneous emission, the appropriate QFI is no longer precisely $4 V(J_y)$, unless every spontaneously emitted particle could be captured and included in the measurement signal. The number of atoms transferred to $\ahat_2$ is approximately $1.6\times10^3$, which roughly corresponds to $N_{a_2}$ at the most sensitive point of Fig.~\ref{pops_noseed}.  For comparison, we have shown the sensitivity in the absence of any temporal shaping ($u_{\mathrm{LO}}(t) \propto 1$), which fails to achieve any significant quantum enhancement. This indicates that the choice of local oscillator is crucial to the ability to extract the best measurement sensitivity from the system. 

\begin{figure}
\includegraphics[width=1\columnwidth]{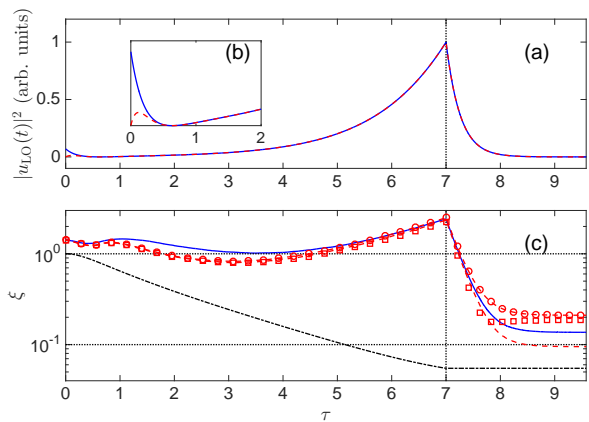}
\caption{(Color online).   Dynamics of a realistic cavity with $\alpha_{20} = -i\sqrt{10^4}$, $\mathcal{C}_{0} = -\sqrt{10^4}$. The atom-light coupling was switched off at $\tau = 7$, as indicated by the vertical black dotted line in both figures. (a) The temporal shaping of $|u_{\mathrm{LO}}(t)|^2$ from \eq{u_opt1}(blue solid line), and \eq{u_opt_seed} (red dashed line) used to calculate $\xi_S$. (b) enlargement to show the details at early times. 
(c) $\xi_S$ using \eq{u_opt1} (blue line), and \eq{u_opt_seed} (red dashed line), and \eq{u_opt_seed} with $1\%$ detection loss of light (red circles) and $1\%$ detection loss of atoms (red squares). For each point in time, a different $u_{\mathrm{LO}}(t)$ was used, such that the normalisation $\int_0^t |u_{\mathrm{LO}}(t^\prime)|^2 dt^\prime = 1$ could be enforced.  Parameters: $\alpha_{10} = \sqrt{N_t-N_{\mathrm{seed}}}$, $\alpha_{20} =-i\sqrt{N_{\mathrm{seed}}}$,  $\mathcal{C}_0 = -\sqrt{N_{\mathrm{seed}}}$,  $N_{\mathrm{seed}} = 10^4$, $N_t = 10^7$, $\beta_0=0$, $\kappa = 13.0 \times 10^6$ rad s$^{-1}$, $\chi = 1.06 \times 10^3$ rad s$^{-1}$, and $\gamma = 600$ rad s$^{-1}$. 
}
\label{delta_phi_realcav2}
\end{figure}

For completeness we have included the case of a coherent seed of $10^4$ particles in both $\ahat_2$ and $\chat$, and chosen the phase such that it corresponds to the initial {\it de-amplification} of atoms (Fig.~\ref{pops_asymseed}). This seed could be created by a coherent Raman transition with classical light, and then allowing the population of cavity photons to decay to the appropriate level before the driving field $\Omega_{13}$ is switched back on. As some of the photons leak out of the cavity during the state preparation stage, the dynamics do not exactly correspond to the case presented in (Fig.~\ref{pops_asymseed}). As such, the QFI does not obtain the same degree of enhancement as the perfect cavity case, but is still better than the case presented in (Fig.~\ref{pops_symseed}). An explanation for this is that as some of the seed leaks out of the cavity, the conditions for perfect de-amplification are no-longer met and the dynamics begins to mimic the case with a large atomic seed and no optical seed. Fig.~\ref{delta_phi_realcav2} shows $\xi_S$, from \eq{sig_cavb}, using \eq{u_opt1}. The sensitivity using this signal is significantly worse than the optimum allowed by the QCRB. By setting the local oscillator intensity proportional to the intensity of $\bhat_{\rm out}$, we are incorporating information from all the photons in the field into our measurement signal. This includes the initial ``seed'' photons, which carry no useful information. We can enhance our signal further by using a different choice of local oscillator which weights the information carried by the photons that arrive later (which are more likely to come from an atom-photon pair creation even) more highly than the early arriving photons (which are more likely to come from the uncorrelated seed). Specifically, we set 
\begin{equation}
u_{\mathrm{LO}}(t) \propto \sqrt{\langle \bhatd_{\rm out} \bhat_{\rm out} \rangle \left(1 - \exp(-\gamma_c t)\right)} \, . \label{u_opt_seed}
\end{equation} 
This function takes into account that the contribution from the seed (which carries no information) will exponentially decay, and that the photons produced during the Raman superradiance process (which carry the correlations with the atomic mode) will begin to dominate on timescales greater than $1/\gamma_c$. Fig.~\ref{delta_phi_realcav2} shows the sensitivity using this modified local oscillator shaping, and demonstrates that it does provide a sensitivity closer to the limit set by the QCRB. 

We have also included the effect of detection inefficiencies, by assuming a simple linear loss model $\hat{q} \rightarrow \sqrt{1-\zeta}\hat{q} + \sqrt{\zeta} \hat{v}$, for $\hat{q} = \{\ahat_1, \ahat_2, \bhat_0\}$ where $\hat{v}$ is assumed to operate on a vacuum mode. As with most schemes that rely on quantum correlations to enhance sensitivity, a small amount of loss can significantly degrade the sensitivity. Interestingly, the performance is more sensitive to loss of photons than loss of atoms. For the parameters used in Fig.~\ref{delta_phi_realcav2}, a detection loss of 1\% of either atoms or photons roughly doubles $\xi$, with loss of atoms performing slightly better than loss of photons. For an atomic loss of 5\%, we found that the sensitivity reached a minimum of $\xi \approx 0.5$, while the the same level of photonic loss reduces the sensitivity to worse than the SQL.

\section{Summary}
In conclusion, we have modelled the generation of atom-light entanglement in an optical cavity via Raman-superradiance, and shown how this entanglement can be used to enhance the sensitivity of atom interferometry. Information from the lighted emitted from the cavity is correlated with the noise in the atom interferometer, so can be used to increase the sensitivity to better than the standard quantum limit. We found that a simple choice of estimator involving the combination of the atomic spin operator $\jhat_z$ and the quadrature of the light was close to optimal, in that it yielded a sensitivity close to the limit set by the QCRB. However, this required carefully selection of the temporal profile of the local-oscillator used to measure the quadrature via homodyne detection. We found that an optical cavity reduces the need for a large atomic seed, which enhances the sensitivity over what we found in \cite{Haine:2013} by a factor of five. We found that for realistic cavity parameters that an enhancement of 50 below the standard quantum limit was achievable.

\section{Acknowledgements}
We would like to acknowledge useful discussions with Stuart Szigeti, Joel Corney, Warwick Bowen, and Samantha Hood. This project was funded by ARC Project No. DE130100575.

\bibliography{simon_bib}

\end{document}